%% file: main.tex
\documentclass[aps,prc,preprint,groupedaddress]{revtex4-2}
\usepackage{graphicx}
\usepackage{bm}
\usepackage{multirow}
\usepackage{amsmath}
\usepackage[
    colorlinks=true, 
    citecolor=blue,  
    urlcolor=blue,   
    linkcolor=blue   
]{hyperref}
\begin{document}


\title{Analysis of the energy and angular distributions of photoneutrons from $^{\rm nat}{\rm Pb}$, $^{197}{\rm Au}$, $^{\rm nat}{\rm Sn}$, $^{\rm nat}{\rm Cu}$, $^{\rm nat}{\rm Fe}$, and $^{\rm nat}{\rm Ti}$ using resonance direct theory}


\author{Hayato Takeshita} \email[]{takeshita\_h@shimz.co.jp}
\author{Kazuaki Kosako}
\author{Norikazu Kinoshita}
\affiliation{Institute of Technology, Shimizu Corporation, Tokyo 135-8530, Japan}
\author{Yukinobu Watanabe}
\affiliation{Department of Advanced Energy Science and Engineering, Kyushu University, Kasuga, Fukuoka 816-8580, Japan}


\date{\today}

\begin{abstract}
Photoneutron double-differential cross sections in the giant dipole resonance (GDR) region were calculated to investigate the underlying nuclear reaction mechanisms, with particular emphasis on the role of the direct process. Contributions from direct, pre-equilibrium, and compound processes were all taken into account. Wilkinson's resonance direct (RD) theory, based on the independent particle model, was applied to describe high-energy neutron emission from the direct process. The angular distribution of neutrons emitted via the RD mechanism was formulated using the Agodi and Courant formalism, which was incorporated into the RD framework. Neutron emission from the pre-equilibrium and compound processes was calculated using the two-component exciton model and the Hauser--Feshbach formalism, respectively. The calculated results were compared with experimental data obtained at NewSUBARU using 16.6-MeV quasi-monochromatic linearly-polarized photon beams. Good agreement between calculations and measurements was observed for Pb, Au, and Sn, confirming the validity of the proposed model. Furthermore, the angular anisotropies of photoneutrons emitted from these elements were investigated, revealing considerable contributions from the RD process at high neutron energies. This study provides a deeper understanding of photoneutron emission mechanisms in the GDR energy region.
\end{abstract}


\maketitle

\section{Introduction \label{sec1}}
Photonuclear reactions provide valuable insights into nuclear theories including nuclear reaction mechanisms, nuclear structures, and nucleon interactions. The primary photoabsorption mechanism in the 10--30 MeV excitation energy range is the giant dipole resonance (GDR), which corresponds to  the collective motion of protons against neutrons. In this energy region, the energy-integrated cross sections nearly exhaust the electric dipole sum rule~\cite{haywardPhotonuclearReactions1970}, indicating that the GDR is primarily governed by E1 transitions.

Prior to the 1950s, only statistical decay processes were considered in photonuclear reactions. Statistical theory predicts a strong suppression of photoproton emission from heavy nuclei owing to the Coulomb barrier. However, experimental results have shown that for nuclei with $A\sim 100$, the ratio of the ($\gamma$,$p$) cross section to the ($\gamma$,$n$) cross section is 20--1000 times larger than predicted by statistical theory~\cite{wafflerRelativeEffectiveCross1948}. Courant~\cite{courantDirectPhotodisiategrationProcesses1951} explained the unexpectedly large photoproton emission cross sections by introducing a direct photoelectric emission process based on an independent particle model. In this model, an incident photon interacts directly with a proton in the nucleus without forming a resonance state, transferring sufficient energy for the proton to overcome the Coulomb barrier. Although the calculations showed that the ($\gamma$,$p$) cross sections increased by a factor of up to approximately 10, the resulting values remained approximately one order of magnitude smaller than the experimental data.

Many authors have extensively measured the energy spectra of protons and neutrons using bremsstrahlung photons with maximum energies below 30 MeV~\cite{zatsepinaANGULARENERGYDISTRIBUTIONS1963,lepestkinEnergyDistributionsPhotoneutrons1985,ishkhanovPhotoprotonEnergySpectra1977,emmaEnergySpectrumPhotoneutrons1961a,mutchlerAngularDistributionsEnergy1966,tomsPhotoprotonsCeBi1953,tomsPhotoprotonsLead208Tantalum1955}. The measured spectra of photoneutrons and photoprotons showed that the high-energy ($\gtrsim$4 MeV) components were considerably larger than predicted by statistical theory, whereas the low-energy ($<$4 MeV) components were in good agreement with theoretical predictions. This discrepancy again highlights the importance of direct processes, in which high-energy nucleons are preferentially emitted.

Statistical theory predicts an isotropic angular distribution of the emitted nucleons. However, experimental data~\cite{mutchlerAngularDistributionsEnergy1966} indicate that high-energy neutrons exhibit anisotropic angular distributions, while low-energy neutrons are emitted isotropically. According to Courant's theory~\cite{courantDirectPhotodisiategrationProcesses1951}, the angular distribution of nucleons emitted via direct processes takes the form $A + B\sin^2\theta$, where the coefficients $A$ and $B$ depend on the angular momentum of the emitted nucleon, and $\theta$ is the angle between the photon and neutron momentum vectors.

Wilkinson introduced the ``resonance direct'' (RD) process for photonuclear reactions, based on the shell model~\cite{wilkinsonNUCLEARPHOTODISINTEGRATION1956}. According to Wilkinson's theory, an incident photon is absorbed by a nucleus, promoting a nucleon from a closed shell to a higher shell that is typically unbound and unoccupied, with a change in orbital angular momentum of one unit (${\it \Delta}l=\pm1$). This transition can involve both valence and core nucleons. The resulting excited nucleus, which corresponds to a one-particle-one-hole (1p1h) state, can subsequently decay via two ways. In the first channel, the nucleon in the unbound state escapes from the nucleus without interacting with the remaining nucleons. In the second channel, the nucleon spreads its energy throughout the nucleus, leading to the formation of a compound nucleus. Wilkinson indicated that the probability of the first process, namely the RD process, is proportional to the ratio of the escape width to the spreading width. Calculations of photoproton spectra from $^{208}{\rm Pb}$ reproduced the experimental data~\cite{tomsPhotoprotonsLead208Tantalum1955} qualitatively using simplified energy-independent optical potential parameters.

Wilkinson's RD theory has been further extended and applied to experimental analyses in subsequent studies. Zatsepina et al.~\cite{zatsepinaANGULARENERGYDISTRIBUTIONS1963} and Mutchler~\cite{mutchlerAngularDistributionsEnergy1966} calculated photoneutron spectra by considering both RD and statistical decay processes. They introduced an energy-dependent optical potential, evaluated by Lane and Wandel~\cite{laneEvaluationImaginaryPart1955}, to calculate the escape width. Angular distributions of directly emitted neutrons were treated using Courant's theory~\cite{courantDirectPhotodisiategrationProcesses1951}. The results successfully reproduced the trends of measured photoneutron spectra from heavy nuclei, but systematically underestimated the magnitude of the direct component. These discrepancies arise from neglecting subsequent interactions between particle and hole states, omission of pre-equilibrium processes, and the use of transmission coefficients instead of penetrability in the evaluation of the spreading width.

In recent years, a research group has systematically measured the double-differential cross sections (DDXs) of photoneutrons in the GDR energy region using linearly-polarized laser Compton scattering (LCS) photons at NewSUBARU~\cite{kiriharaNeutronEmissionSpectrum2020,kimtuyetEnergyAngularDistribution2021,hongthuongExperimentalStudyPhotoneutron2024,thuongPhotoneutronEmissionProcess2025}. The group observed a considerable direct component in the 4--8 MeV region of the DDXs for $^{\rm nat}{\rm Pb}$, $^{197}{\rm Au}$, and $^{\rm nat}{\rm Sn}$ under irradiation with quasi-monochromatic 16.6-MeV LCS photons. Angular differential cross sections (ADXs) were obtained by integrating the DDXs over neutron emission energies above 2 MeV. The ADXs were well fitted by the expression $a + b\cos(2\Theta)$, where $\Theta$ is the angle between the photon polarization direction and the neutron momentum vector. The ratio of the coefficients $b/a$ represents the angular anisotropy of the emitted neutrons, as predicted by Agodi~\cite{agodiGPolarizationEffectsPhotonuclear1957}. Agodi's formula is equivalent to Courant's formula~\cite{courantDirectPhotodisiategrationProcesses1951} when averaged over polarization angles. A recent study by the same group suggested that the ratio $b/a$ near the maximum neutron emission energy can be approximated using Courant's theory~\cite{courantDirectPhotodisiategrationProcesses1951} based on the outermost shell orbital of the target nuclei. However, the energy dependence of the ratio $b/a$ has not yet been discussed.

This study aims to investigate the neutron production mechanism in photonuclear reactions, considering direct, pre-equilibrium, and compound processes in the GDR energy region. The direct process is described based on Wilkinson's RD theory, while the pre-equilibrium and compound processes are calculated using the exciton model and the Hauser--Feshbach formalism, respectively. Angular distributions of directly emitted neutrons are treated using Agodi's and Courant's  theories. The applicability of this model is discussed by comparing calculated results with neutron DDXs for $^{\rm nat}{\rm Pb}$, $^{197}{\rm Au}$, $^{\rm nat}{\rm Sn}$, $^{\rm nat}{\rm Cu}$, $^{\rm nat}{\rm Fe}$, and $^{\rm nat}{\rm Ti}$ measured at NewSUBARU~\cite{kimtuyetEnergyAngularDistribution2021}. Furthermore, it is shown that the angular anisotropies of neutrons emitted from heavy nuclei can be successfully reproduced by the model.

Section~\ref{sec2}, describes the details of our model, including Wilkinson's RD theory, and outlines the method used to calculated the energy and angular distributions of neutrons directly emitted by linearly polarized photons. Section~\ref{sec3} validates our calculations through comparison with NewSUBARU data. Conclusions and future work are presented in Sec.~\ref{sec4}.

\section{Theoretical model\label{sec2}}
\begin{figure}
    \includegraphics[]{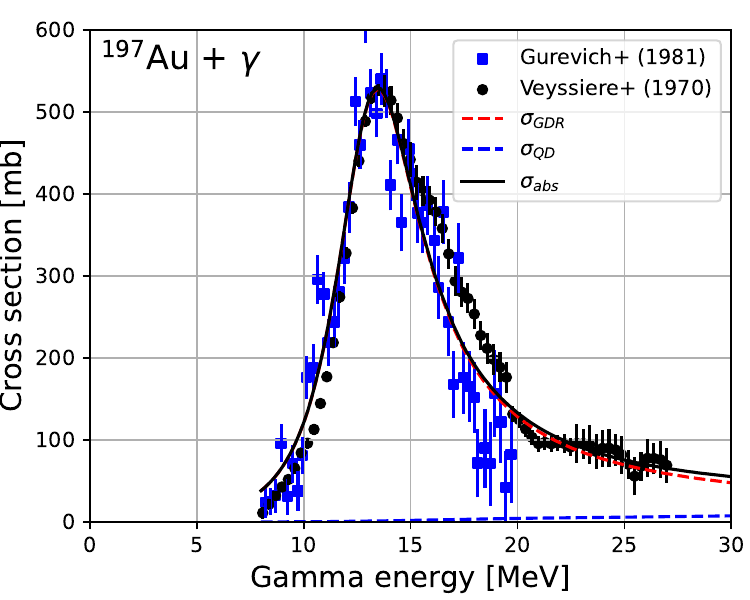}
    \caption{Photoabsorption cross sections for ${}^{197}\text{Au}$. The cross sections for the GDR and QD processes were calculated using TALYS~\cite{koningTALYSModelingNuclear2023}. Experimental data were taken from Refs.~\cite{GUREVICH1981257,VEYSSIERE1970561}.\label{fig1}}
\end{figure}

The photoabsorption cross sections $\sigma_{\rm abs}$ below the pion production threshold energy ($E_\gamma\lesssim$140 MeV) can be expressed as

\begin{equation}
    \label{eq:xs}
    \sigma_{\rm abs} = \sigma_{\rm GDR} + \sigma_{\rm QD},
\end{equation}

\noindent
where $\sigma_{\rm GDR}$ and $\sigma_{\rm QD}$ are the cross sections of the GDR and quasi-deuteron (QD) processes, respectively~\cite{koningTALYSModelingNuclear2023}. Figure~\ref{fig1} shows the measured $\sigma_{\rm abs}$ for $^{197}{\rm Au}$~\cite{GUREVICH1981257,VEYSSIERE1970561}, and the values calculated using TALYS~\cite{koningTALYSModelingNuclear2023}. The red and blue dashed curves represent the calculated $\sigma_{\rm GDR}$ and $\sigma_{\rm QD}$, respectively, while the black solid curve shows their sum. In general, the contribution of the QD process is negligible for incident photon energies $E_\gamma \lesssim 30$ MeV. Therefore, the QD process is neglected hereafter, and the photoabsorption cross section is approximated as $\sigma_{\rm abs} \sim \sigma_{\rm GDR}$.

Following photoabsorption, the excited nucleus decays through a sequence of direct, pre-equilibrium, and compound processes. The remainder of this section describes the mechanisms of neutron emission in each of these processes.

\subsection{Resonance direct process\label{sec2-1}}
To describe the excitation and decay processes, we use Wilkinson's independent particle model~\cite{wilkinsonNUCLEARPHOTODISINTEGRATION1956}. In this framework, an incident photon is absorbed by the nucleus, promoting a nucleon from a closed shell a higher shell with an orbital angular momentum differing by one unit. The transition matrix elements for this process are given by

\begin{equation}
    \label{eq:m2}
    |M_{fi}|^2 = |\langle\psi_f|\hat{D_z}|\psi_i\rangle|^2,
\end{equation}

\noindent
where $|\psi_i\rangle$ and $|\psi_f\rangle$ denote the initial and final single-particle states of the nucleon, respectively, and $\hat{D_z}$ is the $z$-component of the electric dipole operator. Assuming a point-like proton charge distribution, the electric dipole operator $\hat{\bm{D}}$ is given by

\begin{equation}
    \label{eq:dipole}
    \hat{\bm{D}} = e\sum_{p=1}^{Z}\left( \hat{\bm{r}_p} - \hat{\bm{R}}\right),
\end{equation}

\noindent
where $e$ is the elementary charge, $\hat{\bm{r}_p}$ is the proton position vector, and $\hat{\bm{R}}$ is the center-of-mass position of the nucleus. The summation runs over all $Z$ protons in the nucleus. Substituting Eq.~(\ref{eq:dipole}) into Eq.~(\ref{eq:m2}) yields~\cite{haywardPhotonuclearReactions1970}

\begin{equation}
    \label{eq:m2_expression}
    |M_{fi}|^2 = \frac{e_{\rm eff}^2}{3}(2j_i+1)\left(j_i,\left.\frac{1}{2},1,0\right|j_f,\frac{1}{2}\right)^2\left|\int{R_i(r)R_f(r)r^3dr}\right|^2,
\end{equation}

\noindent
where $e_{\rm eff}$ is the effective charge of a nucleon, i.e., $+e(A-Z)/A$ for protons and $-eZ/A$ for neutrons, $A$ is the mass number, $j_i$, $j_f$, $R_i$, and $R_f$ denote the angular momenta and radial wave functions of the initial and final states, respectively. The Clebsch--Gordan coefficients are determined according to the transition type:

\begin{equation}
    \label{eq:matrix_ang}
    (2j_i+1)\left(j_i,\left.\frac{1}{2},1,0\right|j_f,\frac{1}{2}\right)^2 =
    \left\{
        \begin{array}{cl}
            \displaystyle\frac{2(l+1)(l+2)}{2l+3}& \text{for}\ l+\frac{1}{2}\rightleftharpoons l+1+\frac{1}{2}, \\[12pt]
            \displaystyle\frac{2(l+1)}{(2l+3)(2l+1)}& \text{for}\ l+\frac{1}{2}\rightleftharpoons l+1-\frac{1}{2}, \\[12pt]
            \displaystyle\frac{2l(l+1)}{2l+1}& \text{for}\ l-\frac{1}{2}\rightleftharpoons l+1-\frac{1}{2}.
        \end{array}
    \right.
\end{equation}

The forms of the radial wave functions $R_i$ and $R_f$ depend on the choice of nuclear potential. In this work, we adopted an infinite square-well potential to calculate the radial wave functions to maintain consistency with the original framework established by Wilkinson~\cite{wilkinsonNUCLEARPHOTODISINTEGRATION1956}. The overlap integral $I$, corresponding to the final factor in Eq.~(\ref{eq:m2_expression}), is defined as

\begin{equation}
    \label{eq:overlap}
    I = R_N^{-2}\left|\int{R_i(r)R_f(r)r^3dr}\right|^2,
\end{equation}

\noindent
where $R_N=1.3A^{1/3}$ fm is the nuclear radius. The values of $I$ were obtained from numerical calculations and are listed in Table~\ref{tb:overlap}. These values are independent of both the nuclear radius and the mass number. This unique feature provides a versatile and scalable model applicable to various nuclei without the need for element-specific parameter adjustment. The overlap integrals corresponding to the $2l \rightarrow 3l+1$ and $3l \rightarrow 4l+1$ transitions were negligibly small and were therefore excluded.

The relative intensity of the $|\psi_i\rangle \rightarrow |\psi_f\rangle$ transition, $b_{fi}$, is defined as

\begin{equation}
    \label{eq:branch}
    b_{fi} = \frac{|M_{fi}|^2}{\sum_{B_i<E_\gamma}|M_{fi}|^2}.
\end{equation}


\noindent
The summation is taken over all transitions for which the binding energy of the initial orbit $B_i$ is less than the energy of the incident photon. In this work, proton transitions are neglected in the calculation of Eq.~(\ref{eq:branch}); this effect will be discussed in Sec.~\ref{sec3}.

\input{table/tableD.tex}

The probability that a neutron with angular momentum $l$ penetrates its centrifugal barrier is given by~\cite{wilkinsonNUCLEARPHOTODISINTEGRATION1956}

\begin{equation}
    \label{eq:transmission}
        C_l = \frac{2kv_l\hbar^2/MR_N}{2W},
\end{equation}

\noindent
where $k = \sqrt{2ME_n}/\hbar$ is the neutron wave number outside the nucleus, $M$ is the neutron mass, and $W$ is the imaginary part of the neutron optical potential. The neutron kinetic energy is given by $E_n=E_\gamma-B_i$. The approximate expression $W = 0.4E_n + 2.5$ MeV was used~\cite{laneEvaluationImaginaryPart1955}. According to Blatt and Weisskopf~\cite{blattTheoreticalNuclearPhysics2012}, the penetrability $v_l$ is given by

\begin{equation}
    \label{eq:penetrability}
        v_l = \frac{2}{\pi kR_N}\frac{1}{\left[J_{l+1/2}\left(kR_N\right)\right]^2 + \left[N_{l+1/2}\left(kR_N\right)\right]^2},
\end{equation}

\noindent
where $J_p(z)$ and $N_p(z)$ are the Bessel and Neumann functions of order $p$, respectively.

The energy-differential cross section of the RD process is obtained by summing the contributions from all orbits involved in the RD transitions and is given by

\begin{equation}
    \label{eq:de_reso}
    \frac{d\sigma_{RD}}{dE_n} = \sigma_{GDR}\sum_{B_i<E_\gamma}\left[b_{fi}C_l\delta\left(E_n-(E_\gamma-B_i)\right)\right],
\end{equation}

\noindent
where the delta function ensures energy conservation. Integrating Eq.~(\ref{eq:de_reso}) over the neutron energy yields the cross section for the RD process:

\begin{equation}
    \label{eq:sigma_reso}
        \sigma_{RD} = \sigma_{GDR}T_{RD},
\end{equation}

\noindent
where $T_{RD}$ is the probability of proceeding via the RD process and is given by

\begin{equation}
    \label{eq:trans}
        T_{RD} = \sum_{B_i<E_\gamma}b_{fi} C_l.
\end{equation}

After the RD process, the residual nucleus may remain in a low-lying excited state characterized by a 0p1h configuration. From this state, the nucleus may emit additional neutrons through subsequent interactions between particles and holes. However, in this work, the excitation energy of the residual nucleus is relatively low, typically up to approximately 8 MeV. Therefore, the contribution from such secondary or subsequent stages of neutron emission is expected to be small. Moreover, neutrons emitted in these later stages would have low energies, where the compound process contribution becomes dominant. For this reason, the additional contribution from multi-stage neutron emission is neglected.

\indent
Tables~\ref{tb:transition_208Pb}--\ref{tb:transition_48Ti} present the available transitions for the most abundant isotopes considered in this work: $^{208}{\rm Pb}$, $^{197}{\rm Au}$, $^{120}{\rm Sn}$, $^{63}{\rm Cu}$, $^{56}{\rm Fe}$, and $^{48}{\rm Ti}$. Transitions with relative intensities smaller than 0.01 were neglected. The binding energies listed in the first column were obtained from numerical calculations using a diffuse nuclear potential~\cite{rossNucleonEnergyLevels1956}. For neutrons, the radial Schr\"odinger equation to be solved is given by

\begin{equation}
    \begin{split}
        \label{eq:neutron_potential}
        -\frac{\hbar^2}{2M}\frac{1}{{r^2}}\frac{\mathrm{d}}{\mathrm{d}r}\left(r^2\frac{\mathrm{d}R}{\mathrm{d}r}\right)
        + \left(
            -\frac{V_{0n}}{1+\exp\left[\alpha(r-R_N)\right]}
             + \frac{\hbar^2}{2M}\frac{l(l+1)}{r^2} \right. \\
             \left. - \frac{\lambda\hbar^2}{4M^2c^2r}\frac{\alpha V_{0n}\exp\left[\alpha(r-R_N)\right]}{\left\{1+\exp\left[\alpha(r-R_N)\right]\right\}^2}
             \boldsymbol{\sigma}\cdot \mathbf{l}\right)R
        = ER,
    \end{split}
\end{equation}


\noindent
where $V_{0n}$ is the neutron well depth, $\alpha=1.45\ \mathrm{fm}^{-1}$ is the diffuseness parameter, $\lambda=39.5$ is the spin-orbit coupling constant, $c$ is the speed of light, and $E$ is the binding energy of the nucleon. The eigenvalues of the spin-orbit coupling operator $\boldsymbol{\sigma}\cdot \mathbf{l}$ are $l$ and $-(l+1)$ for states with $j=l+\frac{1}{2}$ and $j=l-\frac{1}{2}$, respectively. The parameter $V_{0n}$ was adjusted to reproduce the neutron separation energies listed in the Live Chart of Nuclides~\cite{LivechartTableNuclides}.

\input{table/tablePb.tex}
\input{table/tableAu.tex}
\input{table/tableSn.tex}
\input{table/tableCu.tex}
\input{table/tableFe.tex}
\input{table/tableTi.tex}

\subsection{Angular distribution\label{sec2-2}}
According to Agodi~\cite{agodiGPolarizationEffectsPhotonuclear1957}, the angular distribution of nucleons emitted in the direct process induced by linearly polarized photons is given by

\begin{equation}
    \label{eq:ang_polarized}
    f_l^{\rm linear}(\theta,\phi) \propto a_l + b_l \cos(2\Theta),
\end{equation}

\noindent
where $\theta$ is the angle between the photon and neutron momentum vectors, $\phi$ is the angle between the photon polarization direction and the plane defined by the photon and neutron momentum vectors, and $\Theta$ is the angle between the photon polarization direction and the neutron momentum vector. The quantities $a_l$ and $b_l$ depend solely on the orbital angular momentum before the transition, $l$, and are independent of the emission energy and the target nuclide. The relationships between the vectors and angles are shown in Fig.~\ref{fig2}. The angle $\Theta$ is related to $\theta$ and $\phi$ as follows~\cite{kiriharaNeutronEmissionSpectrum2020}:

\begin{figure}
    \includegraphics[]{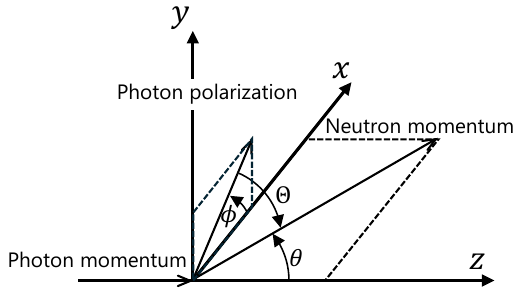}
    \caption{Illustration of the relationship among the angles $\theta$, $\phi$, and $\Theta$.\label{fig2}}
\end{figure}

\begin{equation}
    \label{eq:cos}
    \cos\Theta = \sin\theta\cos\phi.
\end{equation}

\noindent
The validity of Eq.~(\ref{eq:ang_polarized}) has been confirmed experimentally~\cite{hayakawaSpatialAnisotropyNeutrons2016,horikawaNeutronAngularDistribution2014a,kiriharaNeutronEmissionSpectrum2020,kimtuyetEnergyAngularDistribution2021}. If the photon is circularly polarized or unpolarized, the angular distribution should be averaged over $\phi$, yielding

\begin{equation}
    \label{eq:ang_pol_avg}
    \langle f_l^{\rm linear}(\theta,\phi)\rangle_\phi = \frac{1}{\pi}\int_{0}^{\pi}\left[a_l + b_l\cos(2\Theta)\right]d\phi = (a_l-b_l) + b_l\sin^2\theta.
\end{equation}

According to Courant~\cite{courantDirectPhotodisiategrationProcesses1951}, the angular distribution for circularly polarized photons, $f_l^{\rm circular}(\theta)$, is

\begin{equation}
    \label{eq:ang_unpolarized}
    f_l^{\rm circular}(\theta) \propto
    \left\{
        \begin{array}{cl}
            l(l+1) + \displaystyle\frac{1}{2}(l+1)(l+2)\sin^2\theta & \text{for}\ l \rightarrow l+1, \\[12pt]
            l(l+1) + \displaystyle\frac{1}{2}l(l-1)\sin^2\theta & \text{for}\ l \rightarrow l-1.
        \end{array}
    \right.
\end{equation}

\noindent
We assume that Eqs.~(\ref{eq:ang_pol_avg}) and (\ref{eq:ang_unpolarized}) are equivalent. This assumption leads to the ratio $b_l/a_l$:


\begin{equation}
    \label{eq:b_a}
    b_l/a_l = 
    \left\{
        \begin{array}{cl}
            \displaystyle\frac{l+2}{3l+2} & \text{for}\ l \rightarrow l+1, \\[12pt]
            \displaystyle\frac{l-1}{3l+1} & \text{for}\ l \rightarrow l-1.
        \end{array}
    \right.
\end{equation}

\noindent
Normalizing Eq.~(\ref{eq:ang_polarized}) gives

\begin{equation}
    \label{eq:fpol}
    \hat{f}_l^{\rm linear}(\theta,\phi) = \frac{3}{4\pi}\frac{(1-b_l/a_l)+2b_l/a_l\sin^2\theta\cos^2\phi}{3-b_l/a_l}.
\end{equation}

Finally, the neutron DDX of the RD process induced by linearly polarized photons is obtained by incorporating the angular distribution, as expressed in Eq.~(\ref{eq:fpol}), into the energy-differential cross section given by Eq.~(\ref{eq:de_reso}):

\begin{equation}
    \label{eq:ddx_reso}
    \frac{d^2\sigma_{RD}}{dE_nd\Omega_n}(E_n,\theta,\phi;E_\gamma) = \sigma_{GDR}(E_\gamma)\sum_{B_i<E_\gamma}\hat{f}_l^{\rm linear}(\theta,\phi)\left[b_{fi}C_l\delta\left(E_n-(E_\gamma-B_i)\right)\right].
\end{equation}

\subsection{Pre-equilibrium and compound processes\label{sec2-3}}
Contributions from pre-equilibrium and compound processes were calculated using the two-component exciton model~\cite{kalbachTwocomponentExcitonModel1986} and the Hauser--Feshbach formalism~\cite{hauserInelasticScatteringNeutrons1952}, respectively. These calculations were performed with TALYS~\cite{koningTALYSModelingNuclear2023}, which incorporates these models. TALYS was chosen owing to its extensive validation against a wide range of experimental data~\cite{koningGlobalPreequilibriumAnalysis2004}. All adjustable parameters were kept at their default values. Therefore, the total DDXs were obtained by 


\begin{equation}
        \label{eq:ddx_sum}
        \frac{d^2\sigma}{dE_nd\Omega_n} = \frac{d^2\sigma_{RD}}{dE_nd\Omega_n} + \left(1 - T_{RD}\right)\left[\frac{d^2\sigma_{PE}}{dE_nd\Omega_n} + \frac{d^2\sigma_{CN}}{dE_nd\Omega_n}\right],
\end{equation}
    
\noindent
\newline
where $d^2\sigma_{PE}/dE_nd\Omega_n$ and $d^2\sigma_{CN}/dE_nd\Omega_n$ are the neutron DDXs from the pre-equilibrium and compound processes, respectively, calculated using TALYS, and $T_{RD}$ is the probability of the RD process given by Eq.~(\ref{eq:trans}). The factor ($1-T_{RD}$) is included to ensure flux conservation.

\section{Results and Discussion\label{sec3}}
\subsection{\label{sec3-1}Consideration of neutron energy smearing}
The calculation results were compared with experimental data measured at NewSUBARU using quasi-monochromatic 16.6-MeV LCS photons~\cite{kimtuyetEnergyAngularDistribution2021}. For this comparison, the results derived from Eq.~(\ref{eq:ddx_sum}) were smeared using the experimental incident photon energy distribution and the detector energy resolution. The incident photon energy distribution $\Phi(E_\gamma)$~\cite{kiriharaNeutronEmissionSpectrum2020} was taken into account by

\begin{equation}
    \label{eq:ddx_avg}
    \left\langle\frac{d^2\sigma}{dE_nd\Omega_n}\right\rangle = \int_{S_n}^{E_{\gamma \rm{max}}}\frac{d^2\sigma}{dE_nd\Omega_n}(E_n,\theta,\phi;E_\gamma)\Phi(E_\gamma)dE_\gamma \left/ \int_{S_n}^{E_{\gamma \rm{max}}}\Phi(E_\gamma)dE_\gamma\right. ,
\end{equation}

\noindent
where the integration extends from the neutron separation energy of the target nucleus $S_n$ to the maximum photon energy $E_{\gamma \rm{max}}$. The detector energy resolution (linearly interpolated from 5\% at $E_n=2$ MeV to 7.5\% at $E_n=8$ MeV) was incorporated by assuming a Gaussian response function.

\begin{figure}
    \includegraphics[]{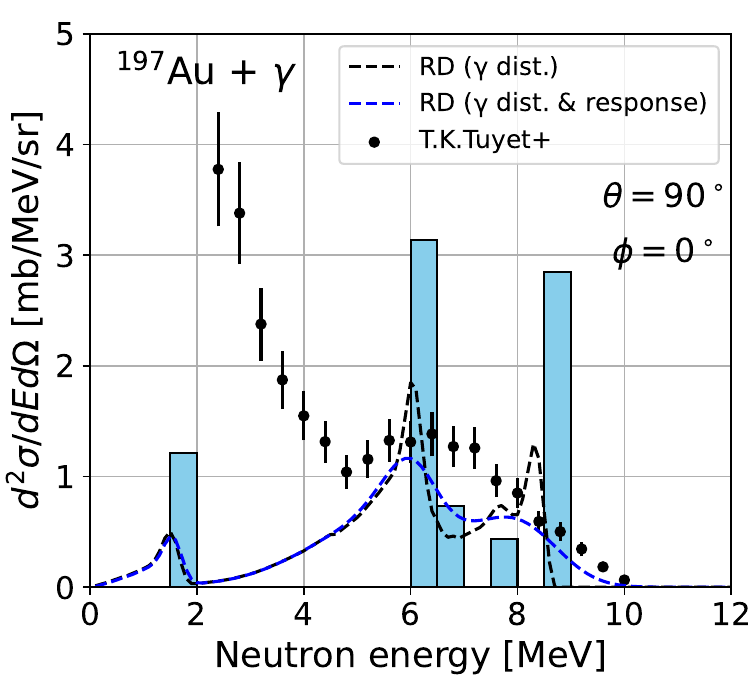}
    \caption{Photoneutron DDXs for $^{197}{\rm Au}$ at an emission angle of $(\theta,\phi) = (90^\circ,0^\circ)$. The histogram represents the theoretical RD component calculated for 16.6-MeV monochromatic photons with a bin width of 0.5 MeV. The black dashed curve shows the RD component smoothed using the incident photon energy distribution. The blue dashed curve shows the RD component smoothed using both the incident photon energy distribution and the detector response.\label{fig:mono}}
\end{figure}

Figure~\ref{fig:mono} shows the calculated DDX for $^{197}{\rm Au}$ at an emission angle of $(\theta,\phi) = (90^\circ,0^\circ)$. The histogram represents the RD component induced by 16.6-MeV monochromatic photons, which was calculated using Eq.~(\ref{eq:ddx_reso}). The black dashed curve shows the DDX smoothed using only the incident photon energy distribution, while the blue dashed curve represents the DDX smoothed using both the incident photon energy distribution and the detector response. To avoid divergences associated with a delta-function representation, the DDX of the RD process prior to smoothing is presented as a histogram with a bin width of 0.5 MeV. The histogram highlights the RD contributions arising from transitions between discrete levels, which are particularly evident for neutron energies above 6 MeV. The energy bins correspond to discrete neutron energies, $E_n = E_\gamma - B_i$, where $B_i$ denote the neutron binding energies listed in Table~\ref{tb:transition_197Au}.

The smoothing process shows that the calculated values tend to converge toward the experimental data. Conversely, this suggests that information on fine shell structures becomes less discernible in the experimental results. However, the  enhancement in the DDX observed above 6 MeV clearly indicates a significant contribution from the RD process.

\subsection{\label{sec3-2}Energy distributions}
\begin{figure*}
    \includegraphics[width=1.0\hsize]{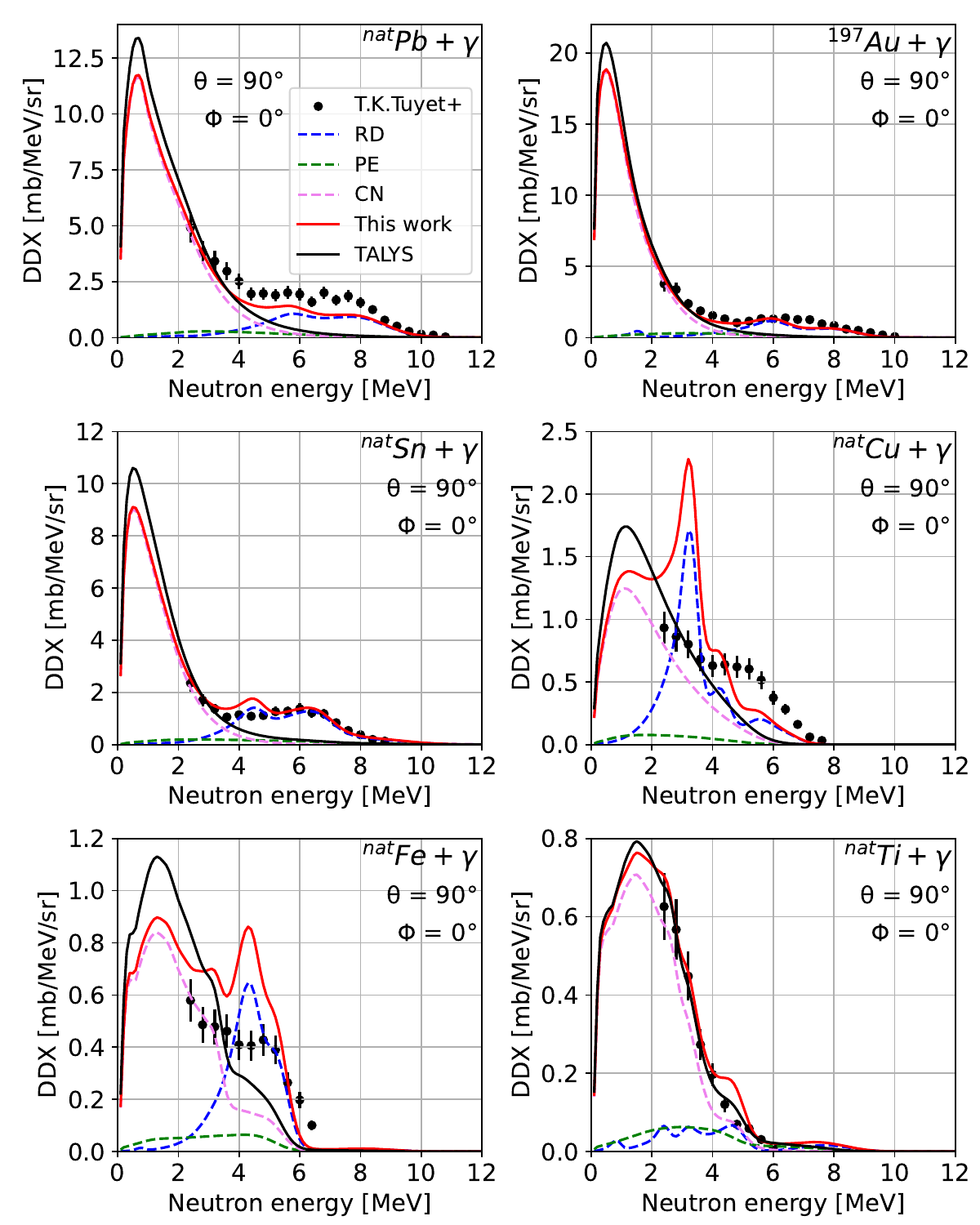}
    \caption{Calculated and experimental DDXs for $^{\rm nat}{\rm Pb}$, $^{197}{\rm Au}$, $^{\rm nat}{\rm Sn}$, $^{\rm nat}{\rm Cu}$, $^{\rm nat}{\rm Fe}$, and $^{\rm nat}{\rm Ti}$ irradiated by quasi-monochromatic 16.6-MeV photons. The data correspond to an emission angle of $(\theta,\phi) = (90^\circ,0^\circ)$.\label{fig:results}}
\end{figure*}

Figure~\ref{fig:results} shows the calculation results for $^{\rm nat}{\rm Pb}$, $^{197}{\rm Au}$, $^{\rm nat}{\rm Sn}$, $^{\rm nat}{\rm Cu}$, $^{\rm nat}{\rm Fe}$, and $^{\rm nat}{\rm Ti}$ at an emission angle of $(\theta,\phi) = (90^\circ,0^\circ)$. For elements other than Au, the DDXs for natural elements were obtained by calculating the DDXs for individual isotopes and summing them, with weighs corresponding to their natural abundances. The calculations for $^{\rm nat}\text{Pb}$, $^{197}\text{Au}$, and $^{\rm nat}\text{Sn}$ show good agreement with the experimental data. The RD process primarily contributes to the neutron emission energy range of 4--8 MeV. However, the RD components overestimate the experimental data for $^{\rm nat}\text{Cu}$, $^{\rm nat}\text{Fe}$, and $^{\rm nat}\text{Ti}$.

The overestimation is attributed to the exclusion of proton transitions in the calculation of Eq.~(\ref{eq:branch}). In the present calculations, we have focused exclusively on the photoneutron emission mechanism, and consequently, proton transition contributions were neglected. A more rigorous treatment would require the inclusion of these proton transition contributions in the denominator of the branching ratio $b_{fi}$ defined in Eq.~(\ref{eq:branch}). It is expected that the exclusion of these channels effectively increases the relative weight of the neutron transitions, contributing to the overestimation of the RD component. For instance, in case of $^{63}\text{Cu}$, only neutrons in the $1f_{5/2}$ or $2p_{3/2}$ orbits can be emitted as detailed in Table~\ref{tb:transition_63Cu}. Among these transitions, the $l+\frac{1}{2} \rightarrow l+1+\frac{1}{2}$ type transitions, namely $1f_{5/2} \rightarrow 1g_{7/2}$ and $2p_{3/2} \rightarrow 2d_{5/2}$ transitions, exhibit high relative intensities. However, neutron emission via the $1f_{5/2} \rightarrow 1g_{7/2}$ transition involves a large angular momentum ($l=4$), leading to a small penetrability $v_l$ as defined in Eq.~(\ref{eq:penetrability}). Therefore, the contribution of the RD process via the $2p_{3/2} \rightarrow 2d_{5/2}$ transition becomes noticeable at $E_\gamma - 13.1\text{ MeV} \approx 3.5\text{ MeV}$.

For $^{56}\text{Fe}$, only transitions from the $2p_{3/2}$ orbit are allowed for $E_\gamma<15.8\text{ MeV}$ (see Table~\ref{tb:transition_56Fe}), resulting in a pronounced peak around 4.5 MeV. Similarly, for $^{48}\text{Ti}$, neutron emission from only the $1f_{7/2}$ orbit is allowed for $E_\gamma<15.6\text{ MeV}$ (see Table~\ref{tb:transition_48Ti}), leading to an overestimation of approximately 4 MeV.

\subsection{\label{sec3-3}Angular distributions}
\begin{figure*}
    \includegraphics[width=1.0\hsize]{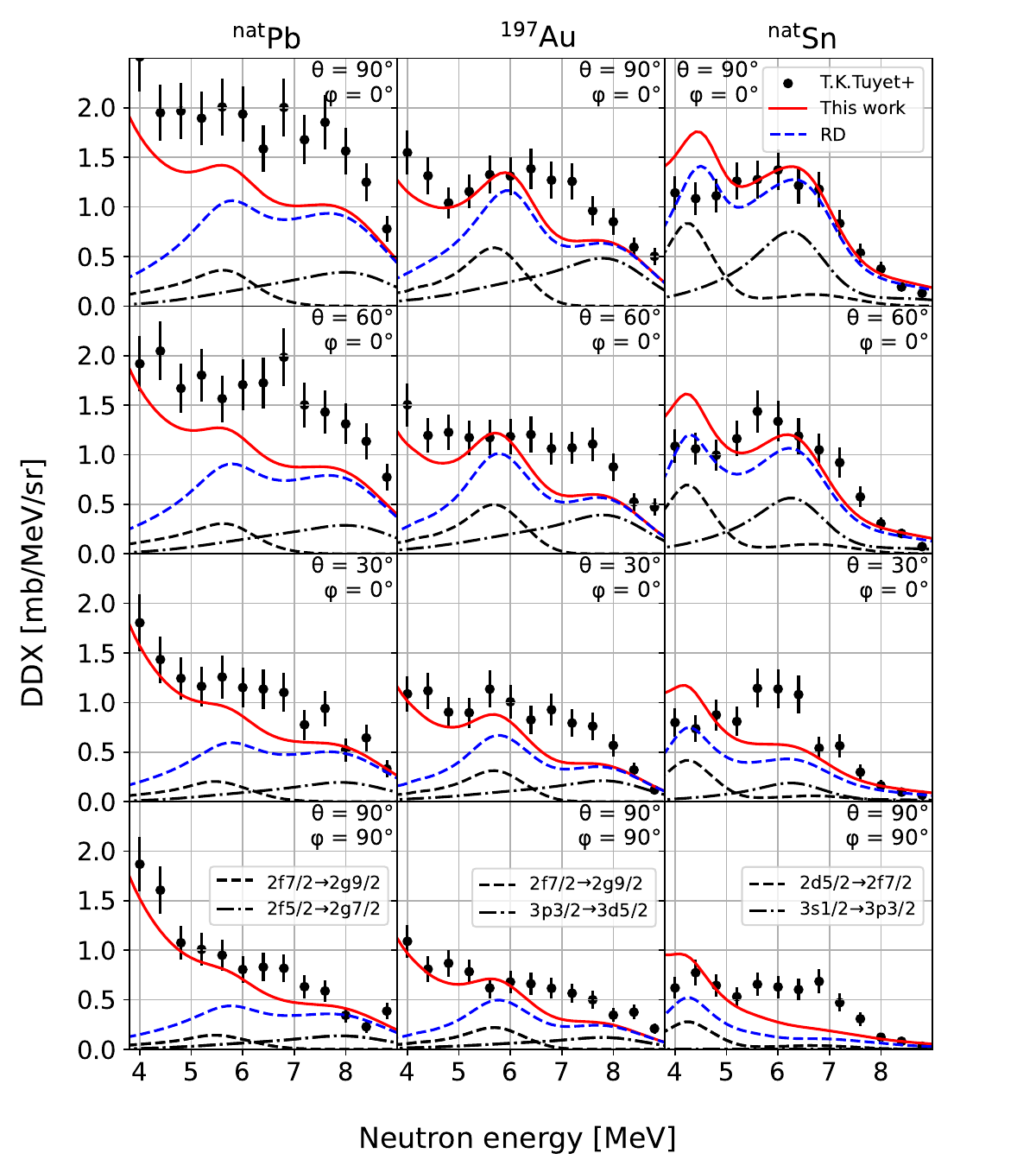}
    \caption{Calculated and experimental DDXs for $^{\rm nat}{\rm Pb}$ (left column), $^{197}{\rm Au}$ (middle column), and $^{\rm nat}{\rm Sn}$ (right column). The panels, arranged from top to bottom, show the DDXs at emission angles ($\theta$, $\phi$) = (90$^\circ$, 0$^\circ$), (60$^\circ$, 0$^\circ$), (30$^\circ$, 0$^\circ$), and (90$^\circ$, 90$^\circ$), which correspond to $\Theta$ = 0$^\circ$, 30$^\circ$, 60$^\circ$, and 90$^\circ$, respectively. The black dashed and dash-dotted curves represent the contributions from the most dominant transitions, where the RD component exhibits two distinct enhancements in the spectra for each element.\label{fig:angle}}
\end{figure*}

Figure~\ref{fig:angle} shows the photoneutron DDXs for $^{\rm nat}{\rm Pb}$, $^{197}{\rm Au}$, and $^{\rm nat}{\rm Sn}$ at emission angles of $(\theta,\phi) = (90^\circ,0^\circ)$, ($60^\circ,0^\circ$), and ($30^\circ,0^\circ$), and ($90^\circ,90^\circ$), which correspond to $\Theta = 0^\circ$, $30^\circ$, $60^\circ$, and $90^\circ$, respectively. These elements were chosen because good agreement between the calculations and experimental values was confirmed, as shown in Fig.~\ref{fig:results}. To focus on the angular dependence of the RD contribution, the figure is restricted to neutron energies above 4 MeV. The black dashed and dash-dotted curves represent the contributions from the most dominant transitions. In these cases, the RD component exhibits two distinct enhancements in the spectra for each element. These transitions are $nl \rightarrow nl+1$ type and have high relative intensities.

\begin{figure}
    \includegraphics[width=8cm]{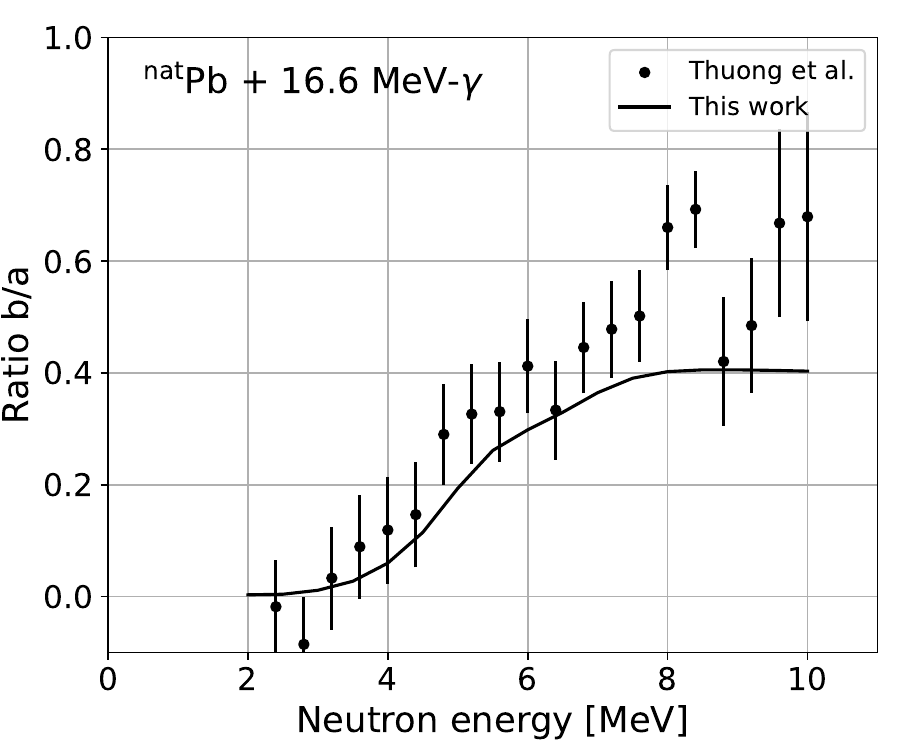}
    \includegraphics[width=8cm]{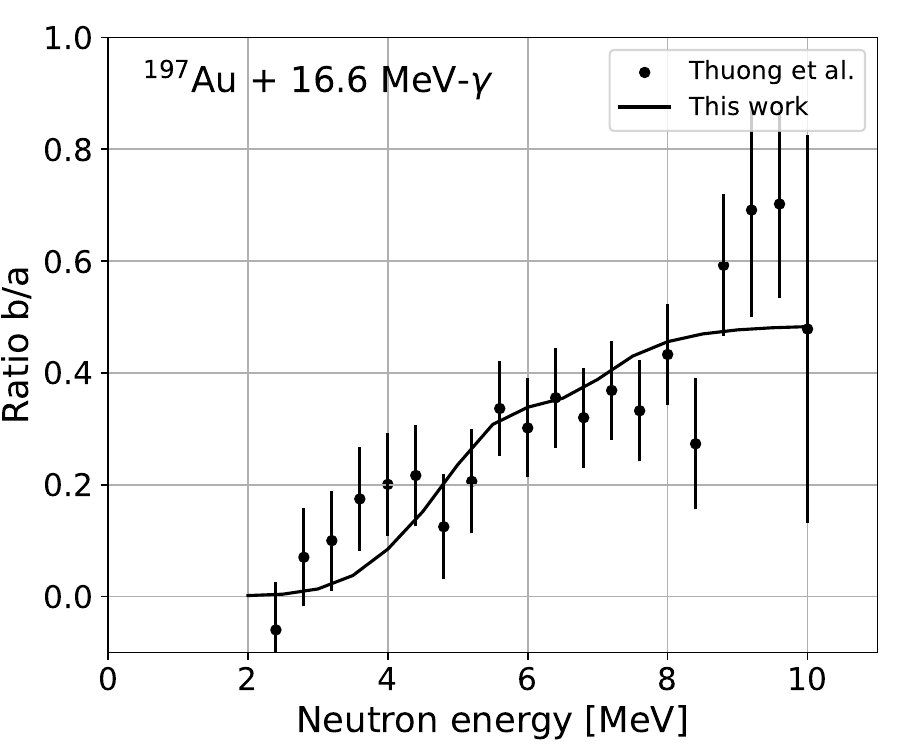}
    \includegraphics[width=8cm]{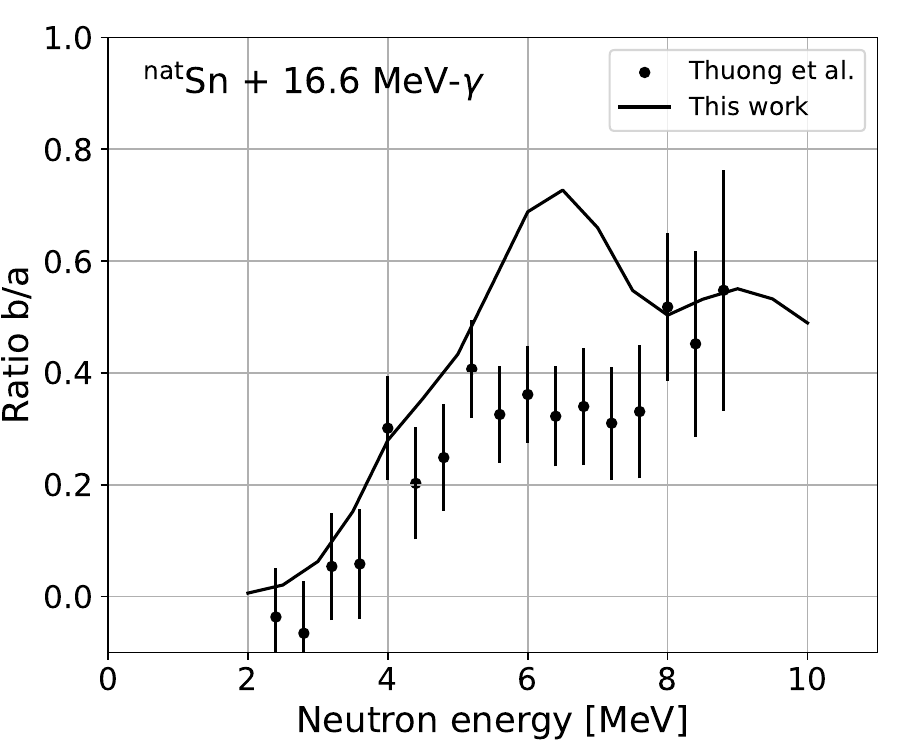}
    \caption{The $b/a$ ratios for the $^{\rm nat}{\rm Pb}$, $^{197}{\rm Au}$, and $^{\rm nat}{\rm Sn}$ targets.\label{fig:ba}}
\end{figure}

To investigate the energy dependence of angular asymmetries, the coefficient ratios $b/a$ were obtained by fitting the DDXs with the function $a+b\cos(2\Theta)$ at each neutron energy bin, following the same procedure as in Ref.~\cite{thuongPhotoneutronEmissionProcess2025}. Figure~\ref{fig:ba} shows the $b/a$ values for $^{\rm nat}{\rm Pb}$, $^{197}{\rm Au}$, and $^{\rm nat}{\rm Sn}$ as a function of neutron energy.

The $b/a$ value increases monotonically with increasing neutron energy except in the high energy region of $^{\rm nat}{\rm Sn}$. In the low-energy region, the $b/a$ value for all three elements is nearly zero. This behavior is attributed to the dominance of the compound process at low energies, as shown in Fig.~\ref{fig:results}. As the neutron energy increases, the contribution of the compound process decreases, while the RD process becomes more significant. This leads to an enhancement of the $b/a$ values.

For $^{\rm nat}{\rm Pb}$, the $2f_{7/2} \rightarrow 2g_{9/2}$ transition is the dominant contributor around 5.5 MeV, as shown in Fig.~\ref{fig:angle}. In the region approximately 8 MeV, the calculated $b/a$ value underestimates the experimental result. As discussed in Ref.~\cite{thuongPhotoneutronEmissionProcess2025}, the $b/a$ value at the maximum emission energy is expected to be approximately 0.6. This expectation is based on the assumption that the outermost shell is the $3p$ orbital, for which the $3p \rightarrow 3d$ transition should dominate. This corresponds to an $l \rightarrow l+1$ transition with $l=1$, as described in Eq.~(\ref{eq:b_a}). However, in our calculation, the $2f_{5/2} \rightarrow 2g_{7/2}$ transition, which corresponds to $b/a=0.45$, is dominant around 8 MeV, and the $b/a$ value remains approximately 0.4.

The tendency for $^{197}{\rm Au}$ is similar to the $^{\rm nat}{\rm Pb}$ case. In particular, for neutron energies lower than 5.5 MeV, the dominant contribution comes from the same $2f_{7/2} \rightarrow 2g_{9/2}$ transition, and therefore the $b/a$ behavior below 5.5 MeV shows a similar tendency. Above this energy, the major contributor is the $3p_{3/2} \rightarrow 3d_{5/2}$ transition, which corresponds to $b/a=0.6$. As a result, the $b/a$ values are slightly larger than in the $^{\rm nat}{\rm Pb}$ case.

For $^{\rm nat}{\rm Sn}$, the calculated $b/a$ value overestimates the experimental values in the 5--8 MeV region. This is because the $3s_{1/2} \rightarrow 3p_{3/2}$ transition, which corresponds to $b/a = 1$, becomes dominant around 6 MeV. However, the experimental $b/a$ value in this energy region is approximately 0.4. This suggests that the contribution from $d\rightarrow f$ transitions is more significant than predicted by the calculation.

\section{Conclusions \label{sec4}}
We proposed a photonuclear reaction model that includes direct, pre-equilibrium, and compound processes. Wilkinson's RD theory was applied to calculate the direct component of the photoneutron spectra. The angular distribution of directly emitted neutrons was determined based on Agodi's and Courant's theories. Our model successfully reproduces the neutron DDXs for $^{\rm nat}{\rm Pb}$, $^{197}{\rm Au}$, and $^{\rm nat}{\rm Sn}$ measured at NewSUBARU. It was determined that the RD component is responsible for neutron emission at energies above 4 MeV. However, the model calculations overestimate the neutron DDXs for $^{\rm nat}{\rm Cu}$, $^{\rm nat}{\rm Fe}$, and $^{\rm nat}{\rm Ti}$. This overestimation is primarily attributed to the exclusion of proton transitions in the calculation of the transition matrix elements. In addition, only a limited number of transitions are allowed for medium-heavy nuclei, which leads to a relatively high intensity for each individual transition. As a result, the calculated neutron DDXs are systematically overestimated.

Angular distributions of emitted neutrons from $^{\rm nat}{\rm Pb}$, $^{197}{\rm Au}$, and $^{\rm nat}{\rm Sn}$ were also investigated. The high-energy component of both the experimental and calculated DDXs decreases with increasing $\Theta$, where $\Theta$ is the angle between the photon polarization direction and the neutron momentum. This behavior follows the $a + b\cos(2\Theta)$ form predicted by Agodi's and Courant's theories. The angular anisotropy of neutrons was quantified by the coefficient ratio $b/a$. While the calculations generally reproduce the energy dependence of $b/a$, some discrepancies are observed for neutron energies above 4 MeV.

Several physical effects were not included in this work. These omissions include the level density of excited states and neutron energy broadening owing to the finite lifetime of excited states in the residual nucleus. Additionally, while the present model focuses exclusively on the photoneutron emission mechanism, a more comprehensive description of photonuclear reactions requires the inclusion of proton transition channels. Although the current approximation is sufficient for the analysis of photoneutron DDXs, future work will involve incorporating proton emission by calculating the corresponding transition matrix elements and applying more rigorous treatments of the Coulomb barrier penetration. Incorporating these effects is expected to improve the predictive capability of the RD theory, even for medium-heavy nuclei ($A<100$).

\section*{Acknowledgements}
The authors are grateful to F.~Minato for valuable discussions and insightful advice concerning the calculation methodology and the physical consistency of our model.

\bibliography{Photonuclear}

\end{document}

%% file: table/tableD.tex
\begin{table}[bt]
     \caption{Overlap integrals $I$ calculated for an infinite well potential \label{tb:overlap}}
     \centering
     \begin{tabular}{|c|c|c|c|c|c|c|c|}
     \hline
         \multirow{2}{*}{Transition type} & \multicolumn{7}{|c|}{$l$} \\
         \cline{2-8}
                  ~ & 0 & 1 & 2 & 3 & 4 & 5 & 6 \\ \hline
         $1l\rightarrow 1l+1$ & 0.28 & 0.38 & 0.44 & 0.49 & 0.53 & 0.56 & 0.58 \\ \hline
         $2l\rightarrow 2l+1$ & 0.23 & 0.28 & 0.33 & 0.37 & 0.39 & 0.42 & 0.44 \\ \hline
         $3l\rightarrow 3l+1$ & 0.22 & 0.25 & 0.28 & - & - & - & -  \\ \hline
         $1l\rightarrow 2l-1$ & - & 0.092 & 0.065 & 0.050 & 0.039 & 0.036 & 0.027 \\ \hline
         $1l\rightarrow 2l+1$ & 0.001 & 0.002 & 0.002 & 0.003 & 0.002 & 0.002 & 0.002 \\ \hline
         $2l\rightarrow 3l-1$ & ~ & 0.12 & 0.09 & 0.07 & 0.06 & 0.05 & 0.05 \\ \hline
         $3l\rightarrow 4l-1$ & - & 0.13 & 0.11 & 0.09 & - & - & -  \\ \hline
     \end{tabular}
 \end{table}

%% file: table/tablePb.tex
\begin{table}[]
    \caption{Available transitions in ${}^{208}\mathrm{Pb}$ for $E_\gamma<17.5$ MeV. Transitions with relative intensities lower than 0.01 were excluded.\label{tb:transition_208Pb}}
    \begin{ruledtabular}
        \begin{tabular}{lll}
        \textrm{Binding energy [MeV]}&
        \textrm{Transition}&
        \textrm{Relative intensity}\\
        \colrule
        7.34 & $3p_{1/2}\ \rightarrow 3d_{3/2}$  & 0.17\\
             & $\ \ \ \ \ \ \ \  \rightarrow 4s_{1/2}$  & 0.04 \\\hline
        7.96 & $2f_{5/2}\ \rightarrow 2g_{7/2}$ & 0.63 \\
             & $\ \ \ \ \ \ \ \  \rightarrow 3d_{5/2}$ & 0.09 \\\hline 
        8.36 & $3p_{3/2}\ \rightarrow 3d_{3/2}$  & 0.03\\
             & $\ \ \ \ \ \ \ \  \rightarrow 3d_{5/2}$  & 0.30 \\
             & $\ \ \ \ \ \ \ \  \rightarrow 4s_{1/2}$  & 0.09 \\\hline
        9.10 & $1i_{13/2}  \rightarrow 1j_{13/2}$ & 0.02 \\
             & $\ \ \ \ \ \ \ \  \rightarrow 1j_{15/2}$ & 2.17 \\
             & $\ \ \ \ \ \ \ \  \rightarrow 2h_{11/2}$  & 0.09 \\\hline
        10.6 & $1h_{9/2}  \rightarrow 1i_{11/2}$  & 1.53 \\
             & $\ \ \ \ \ \ \ \  \rightarrow 2g_{7/2}$  & 0.08 \\\hline
        10.6 & $2f_{7/2}  \rightarrow 2g_{7/2}$  & 0.02 \\
             & $\ \ \ \ \ \ \ \  \rightarrow 2g_{9/2}$ & 0.82 \\
             & $\ \ \ \ \ \ \ \  \rightarrow 3d_{5/2}$  & 0.13 \\\hline
        15.1 & $1h_{11/2} \rightarrow 1i_{11/2}$  & 0.02 \\
             & $\ \ \ \ \ \ \ \  \rightarrow 2g_{9/2}$  & 0.10 \\
    \end{tabular}
        \end{ruledtabular}
    \end{table}

%% file: table/tableAu.tex
\begin{table}[]
    \caption{Same as TABLE~\ref{tb:transition_208Pb}, but for ${}^{197}$Au\label{tb:transition_197Au}}
    \begin{ruledtabular}
    \begin{tabular}{lll}
    \textrm{Binding energy [MeV]}&
    \textrm{Transition}&
    \textrm{Relative intensity}\\
    \colrule
    8.07 & $3p_{3/2}\ \rightarrow 3d_{3/2}$  & 0.03\\
         & $\ \ \ \ \ \ \ \  \rightarrow 3d_{5/2}$  & 0.30\\
         & $\ \ \ \ \ \ \ \  \rightarrow 4s_{1/2}$  & 0.09 \\\hline
    8.77 & $1i_{13/2} \rightarrow 1j_{13/2}$ & 0.02 \\
         & $\ \ \ \ \ \ \ \  \rightarrow 1j_{15/2}$  & 2.17 \\
         & $\ \ \ \ \ \ \ \  \rightarrow 2h_{11/2}$ & 0.09 \\\hline 
    10.1 & $1h_{9/2}  \rightarrow 1i_{11/2}$ & 1.53 \\
         & $\ \ \ \ \ \ \ \  \rightarrow 2g_{7/2}$  & 0.08 \\\hline
    10.3 & $2f_{7/2}  \rightarrow 2g_{7/2}$  & 0.02 \\
         & $\ \ \ \ \ \ \ \  \rightarrow 2g_{9/2}$  & 0.82 \\
         & $\ \ \ \ \ \ \ \  \rightarrow 3d_{5/2}$  & 0.13 \\\hline
    14.9 & $3s_{1/2}  \rightarrow 3p_{1/2}$  & 0.07 \\\hline
    14.9 & $1h_{11/2} \rightarrow 1i_{11/2}$  & 0.02 \\
         & $\ \ \ \ \ \ \ \  \rightarrow 2g_{9/2}$  & 0.10 \\\hline 
    15.0 & $2d_{3/2}  \rightarrow 2f_{5/2}$  & 0.40 \\
         & $\ \ \ \ \ \ \ \  \rightarrow 3p_{1/2}$  & 0.06 \\\hline 
    16.7 & $2d_{5/2}  \rightarrow 2f_{5/2}$  & 0.03 \\\hline
    17.4 & $1g_{7/2}  \rightarrow 2f_{5/2}$ & 0.07 \\
\end{tabular}
    \end{ruledtabular}
\end{table}

%% file: table/tableSn.tex
\begin{table}[]
    \caption{Same as TABLE~\ref{tb:transition_208Pb}, but for ${}^{120}$Sn\label{tb:transition_120Sn}}
    \begin{ruledtabular}
    \begin{tabular}{lll}
    \textrm{Binding energy [MeV]}&
    \textrm{Transition}&
    \textrm{Relative intensity}\\
    \colrule
    9.11 & $1h_{11/2}\ \rightarrow 1i_{13/2}$  & 0.60\\
         & $\ \ \ \ \ \ \ \        \rightarrow 2g_{9/2}$  & 0.03 \\\hline
    9.65 & $3s_{1/2}  \rightarrow 3p_{1/2}$ & 0.07 \\
         & $\ \ \ \ \ \ \ \        \rightarrow 3p_{3/2}$ & 0.15 \\
    10.7 & $1g_{7/2}\ \rightarrow 1h_{9/2}$  & 1.18\\
         & $\ \ \ \ \ \ \ \        \rightarrow 2f_{5/2}$  & 0.07 \\\hline
    11.7 & $2d_{5/2}\ \rightarrow 2f_{5/2}$  & 0.03\\
         & $\ \ \ \ \ \ \ \        \rightarrow 2f_{7/2}$  & 0.57 \\
         & $\ \ \ \ \ \ \ \        \rightarrow 3p_{3/2}$  & 0.11 \\
\end{tabular}
    \end{ruledtabular}
\end{table}

%% file: table/tableCu.tex
\begin{table}[]
    \caption{Same as TABLE~\ref{tb:transition_208Pb}, but for ${}^{63}$Cu\label{tb:transition_63Cu}}
    \begin{ruledtabular}
    \begin{tabular}{lll}
    \textrm{Binding energy [MeV]}&
    \textrm{Transition}&
    \textrm{Relative intensity}\\
    \colrule
    10.8 & $1f_{5/2}\ \rightarrow 1g_{7/2}$  & 0.28\\
    & $\ \ \ \ \ \ \ \        \rightarrow 2d_{3/2}$  & 0.02 \\\hline
    13.1 & $2p_{3/2}  \rightarrow 2d_{3/2}$ & 0.04 \\
         & $\ \ \ \ \ \ \ \        \rightarrow 2d_{5/2}$ & 0.34 \\
         & $\ \ \ \ \ \ \ \        \rightarrow 3s_{1/2}$ & 0.08 \\ 
\end{tabular}
    \end{ruledtabular}
\end{table}

%% file: table/tableFe.tex
\begin{table}[]
    \caption{Same as TABLE~\ref{tb:transition_208Pb}, but for ${}^{56}$Fe\label{tb:transition_56Fe}}
    \begin{ruledtabular}
    \begin{tabular}{lll}
    \textrm{Binding energy [MeV]}&
    \textrm{Transition}&
    \textrm{Relative intensity}\\
    \colrule
    11.2 & $2p_{3/2}\ \rightarrow 2d_{3/2}$  & 0.02\\
    & $\ \ \ \ \ \ \ \        \rightarrow 2d_{5/2}$  & 0.17 \\
    & $\ \ \ \ \ \ \ \        \rightarrow 3s_{1/2}$  & 0.04 \\\hline
    15.8 & $1f_{7/2}  \rightarrow 1g_{7/2}$ & 0.03 \\
         & $\ \ \ \ \ \ \ \        \rightarrow 1g_{9/2}$ & 1.09 \\
         & $\ \ \ \ \ \ \ \        \rightarrow 2d_{5/2}$ & 0.09 \\
\end{tabular}
    \end{ruledtabular}
\end{table}

%% file: table/tableTi.tex
\begin{table}[]
    \caption{Same as TABLE~\ref{tb:transition_208Pb}, but for ${}^{48}$Ti\label{tb:transition_48Ti}}
    \begin{ruledtabular}
    \begin{tabular}{lll}
    \textrm{Binding energy [MeV]}&
    \textrm{Transition}&
    \textrm{Relative intensity}\\
    \colrule
    11.7 & $1f_{7/2}\ \rightarrow 1g_{7/2}$  & 0.02\\
         & $\ \ \ \ \ \ \ \        \rightarrow 1g_{9/2}$  & 0.82 \\
         & $\ \ \ \ \ \ \ \        \rightarrow 2d_{5/2}$  & 0.06 \\\hline
    15.6 & $1d_{3/2}  \rightarrow 1f_{5/2}$ & 0.53 \\
         & $\ \ \ \ \ \ \ \        \rightarrow 2p_{1/2}$ & 0.04 \\ \hline
    16.8 & $2s_{1/2}\ \rightarrow 2p_{1/2}$  & 0.08\\
         & $\ \ \ \ \ \ \ \        \rightarrow 2p_{3/2}$  & 0.15 \\
\end{tabular}
    \end{ruledtabular}
\end{table}

%% file: Photonuclear.bib
@article{agodiGPolarizationEffectsPhotonuclear1957,
  title = {On {$\gamma$}-{{Polarization Effects}} in {{Photonuclear Reactions}}.},
  author = {Agodi, A},
  year = 1957,
  journal = {NUEVO CIMENTO},
  volume = {5},
  number = {1},
 url = {https://doi.org/10.1007/BF02812814}
}

@book{blattTheoreticalNuclearPhysics2012,
  title = {Theoretical Nuclear Physics},
  author = {Blatt, John Markus and Weisskopf, Victor Frederick},
  year = 1979,
  publisher = {Springer New York, NY},
  urldate = {2025-11-11},
  doi = {https://doi.org/10.1007/978-1-4612-9959-2},
  url = {https://doi.org/10.1007/978-1-4612-9959-2}
}

@article{courantDirectPhotodisiategrationProcesses1951,
  title = {Direct Photodisintegration Processes in Nuclei},
  author = {Courant, Ernest D.},
  journal = {Phys. Rev.},
  volume = {82},
  issue = {5},
  pages = {703--709},
  numpages = {0},
  year = {1951},
  month = {Jun},
  publisher = {American Physical Society},
  doi = {10.1103/PhysRev.82.703},
  url = {https://link.aps.org/doi/10.1103/PhysRev.82.703}
}

@article{emmaEnergySpectrumPhotoneutrons1961a,
  title = {Energy Spectrum of Photoneutrons from Cobalt},
  author = {Emma, V. and Milone, C. and Rubbino, A. and Jannelli, S. and Mezzawares, P.},
  year = 1961,
  month = oct,
  journal = {Il Nuovo Cimento},
  volume = {22},
  number = {1},
  pages = {135--144},
  issn = {0029-6341, 1827-6121},
  urldate = {2025-12-12},
  url = {https://doi.org/10.1007/BF02829000},
  copyright = {http://www.springer.com/tdm},
  langid = {english}
}

@article{GUREVICH1981257,
  title = {Total Nuclear Photoabsorption Cross Sections in the Region 150 {$<$} {{A}} {$<$} 190},
  author = {Gurevich, G.M. and Lazareva, L.E. and Mazur, V.M. and Merkulov, {\relax S.YU}. and Solodukhov, G.V. and Tyutin, V.A.},
  year = 1981,
  journal = {Nuclear Physics A},
  volume = {351},
  number = {2},
  pages = {257--268},
  issn = {0375-9474},
  url = {https://doi.org/10.1016/0375-9474(81)90443-7}
}

@article{hauserInelasticScatteringNeutrons1952,
  title = {The Inelastic Scattering of Neutrons},
  author = {Hauser, Walter and Feshbach, Herman},
  journal = {Phys. Rev.},
  volume = {87},
  issue = {2},
  pages = {366--373},
  numpages = {0},
  year = {1952},
  month = {Jul},
  publisher = {American Physical Society},
  doi = {10.1103/PhysRev.87.366},
  url = {https://link.aps.org/doi/10.1103/PhysRev.87.366}
}

@article{hayakawaSpatialAnisotropyNeutrons2016,
  title = {Spatial anisotropy of neutrons emitted from the $^{56}\mathrm{Fe}(\ensuremath{\gamma},\phantom{\rule{0.16em}{0ex}}n)^{55}\mathrm{Fe}$ reaction with a linearly polarized $\ensuremath{\gamma}$-ray beam},
  author = {Hayakawa, T. and Shizuma, T. and Miyamoto, S. and Amano, S. and Takemoto, A. and Yamaguchi, M. and Horikawa, K. and Akimune, H. and Chiba, S. and Ogata, K. and Fujiwara, M.},
  journal = {Phys. Rev. C},
  volume = {93},
  issue = {4},
  pages = {044313},
  numpages = {4},
  year = {2016},
  month = {Apr},
  publisher = {American Physical Society},
  doi = {10.1103/PhysRevC.93.044313},
  url = {https://link.aps.org/doi/10.1103/PhysRevC.93.044313}
}

@book{haywardPhotonuclearReactions1970,
  title={Photonuclear Reactions},
  author={Hayward, E.},
  lccn={70606490},
  series={NBS monograph},
  url={https://books.google.co.jp/books?id=9m3SN33iXvkC},
  year={1970},
  publisher={U.S. National Bureau of Standards}
}

@article{hongthuongExperimentalStudyPhotoneutron2024,
  title = {Experimental Study of Photoneutron Spectra from Tantalum, Tungsten, and Bismuth Targets for 16.6 {{MeV}} Polarized Photons},
  author = {Hong Thuong, Nguyen Thi and Sanami, Toshiya and Yamazaki, Hirohito and Itoga, Toshiro and Kirihara, Yoichi and Sugihara, Kenta and Tuyet, Tran Kim and Mohd Zin, Mohd Faiz Bin and Miyamoto, Shuji and Hashimoto, Satoshi and Asano, Yoshihiro},
  year = 2024,
  month = feb,
  journal = {Journal of Nuclear Science and Technology},
  volume = {61},
  number = {2},
  pages = {261--268},
  issn = {0022-3131, 1881-1248},
  urldate = {2025-11-06},
  url = {https://doi.org/10.1080/00223131.2023.2295438}
}

@article{horikawaNeutronAngularDistribution2014a,
  title = {Neutron Angular Distribution in ({$\gamma$},n) Reactions with Linearly Polarized {$\gamma$} -Ray Beam Generated by Laser {{Compton}} Scattering},
  author = {Horikawa, K. and Miyamoto, S. and Mochizuki, T. and Amano, S. and Li, D. and Imasaki, K. and Izawa, Y. and Ogata, K. and Chiba, S. and Hayakawa, T.},
  year = 2014,
  month = oct,
  journal = {Physics Letters B},
  volume = {737},
  pages = {109--113},
  issn = {03702693},
  urldate = {2025-10-30},
  langid = {english},
  url = {https://doi.org/10.1016/j.physletb.2014.08.024.}
}

@article{ishkhanovPhotoprotonEnergySpectra1977,
  title = {Photoproton Energy Spectra and Isospin Effects in the Decay of Highly Excited States of {{Ni}} Isotopes},
  author = {Ishkhanov, B. S. and Kapitonov, I. M. and Shevchenko, V. G. and Shvedunov, V. I. and Varlamov, V. V.},
  year = 1977,
  month = jun,
  journal = {Nuclear Physics A},
  volume = {283},
  number = {2},
  pages = {307--325},
  issn = {0375-9474},
  urldate = {2025-12-12},
  url = {https://doi.org/10.1016/0375-9474(77)90433-X}
}

@article{kalbachTwocomponentExcitonModel1986,
  title = {Two-component exciton model: Basic formalism away from shell closures},
  author = {Kalbach, C.},
  journal = {Phys. Rev. C},
  volume = {33},
  issue = {3},
  pages = {818--833},
  numpages = {0},
  year = {1986},
  month = {Mar},
  publisher = {American Physical Society},
  doi = {10.1103/PhysRevC.33.818},
  url = {https://link.aps.org/doi/10.1103/PhysRevC.33.818}
}

@article{kimtuyetEnergyAngularDistribution2021,
  title = {Energy and Angular Distribution of Photo-Neutrons for 16.6 {{MeV}} Polarized Photon on Medium--Heavy Targets},
  author = {Kim Tuyet, Tran and Sanami, Toshiya and Yamazaki, Hirohito and Itoga, Toshiro and Takeuchi, Akihiro and Namito, Yoshihito and Miyamoto, Shuji and Asano, Yoshihiro},
  year = 2021,
  month = feb,
  journal = {Nuclear Instruments and Methods in Physics Research Section A: Accelerators, Spectrometers, Detectors and Associated Equipment},
  volume = {989},
  pages = {164965},
  publisher = {Elsevier BV},
  issn = {0168-9002},
  urldate = {2025-07-25},
  copyright = {https://www.elsevier.com/tdm/userlicense/1.0/},
  url = {https://doi.org/10.1016/j.nima.2020.164965.}
}

@article{kiriharaNeutronEmissionSpectrum2020,
  author = {Yoichi Kirihara and Hiroshi Nakashima and Toshiya Sanami and Yoshihito Namito and Toshiro Itoga and Shuji Miyamoto and Akinori Takemoto and Masashi Yamaguchi and Yoshihiro Asano},
  title = {Neutron emission spectrum from gold excited with 16.6 MeV linearly polarized monoenergetic photons},
  journal = {Journal of Nuclear Science and Technology},
  volume = {57},
  number = {4},
  pages = {444--456},
  year = {2020},
  publisher = {Taylor \& Francis},
  doi = {10.1080/00223131.2019.1691073},
  url = {https://doi.org/10.1080/00223131.2019.1691073},
}

@article{koningGlobalPreequilibriumAnalysis2004,
  title = {A Global Pre-Equilibrium Analysis from 7 to 200 {{MeV}} Based on the Optical Model Potential},
  author = {Koning, A.J. and Duijvestijn, M.C.},
  year = 2004,
  month = nov,
  journal = {Nuclear Physics A},
  volume = {744},
  pages = {15--76},
  issn = {03759474},
  urldate = {2025-12-15},
  copyright = {https://www.elsevier.com/tdm/userlicense/1.0/},
  url = {https://doi.org/10.1016/j.nuclphysa.2004.08.013.}
}

@article{koningTALYSModelingNuclear2023,
  title = {{{TALYS}}: Modeling of Nuclear Reactions},
  shorttitle = {{{TALYS}}},
  author = {Koning, Arjan and Hilaire, Stephane and Goriely, Stephane},
  year = 2023,
  month = jun,
  journal = {The European Physical Journal A},
  volume = {59},
  number = {6},
  pages = {131},
  issn = {1434-601X},
  urldate = {2025-12-15},
  url = {https://doi.org/10.1140/epja/s10050-023-01034-3}
}

@article{laneEvaluationImaginaryPart1955,
 author = {Lane, A. M. and Wandel, C. F.},
  journal = {Phys. Rev.},
  volume = {98},
  issue = {5},
  pages = {1524--1525},
  numpages = {0},
  year = {1955},
  month = {Jun},
  publisher = {American Physical Society},
  doi = {10.1103/PhysRev.98.1524},
  url = {https://link.aps.org/doi/10.1103/PhysRev.98.1524}
  }

@article{lepestkinEnergyDistributionsPhotoneutrons1985,
  title = {Energy Distributions of Photoneutrons from Heavy Nuclei at {$E_{\gamma max} = 28.5$} {{MeV}}},
  author = {Lepestkin, A. I. and Seliverstov, V. A. and Sidorov, V. I.},
  year = 1985,
  journal = {Sov. J. Nucl. Phys.(Engl. Transl.);(United States)},
  volume = {42},
  number = {4},
  publisher = {Saratov State University},
  urldate = {2025-12-12}
}

@misc{LivechartTableNuclides,
  title = {Livechart - {{Table}} of {{Nuclides}} - {{Nuclear}} Structure and Decay Data},
  urldate = {2025-11-11},
  url = {https://www-nds.iaea.org/relnsd/vcharthtml/VChartHTML.html}
}

@phdthesis{mutchlerAngularDistributionsEnergy1966,
  title = {The {{Angular Distributions}} and {{Energy Spectra}} of {{Photoncutrons}} from {{Heavy Elements}}},
  author = {Mutchler, G. S.},
  year = 1966,
  school = {Massachusetts institute of technology},
  url = {http://hdl.handle.net/1721.1/13392}
}

@article{rossNucleonEnergyLevels1956,
  title = {Nucleon Energy Levels in a Diffuse Potential},
  author = {Ross, A. A. and Mark, Hans and Lawson, R. D.},
  journal = {Phys. Rev.},
  volume = {102},
  issue = {6},
  pages = {1613--1620},
  numpages = {0},
  year = {1956},
  month = {Jun},
  publisher = {American Physical Society},
  doi = {10.1103/PhysRev.102.1613},
  url = {https://link.aps.org/doi/10.1103/PhysRev.102.1613}
}

@article{thuongPhotoneutronEmissionProcess2025,
  title = {Photoneutron Emission Process on Nuclei around {{A}} = 200 for Giant Dipole Resonance Energies Based on Neutron Energy and Angular Distribution},
  author = {Thuong, Nguyen Thi Hong and Sanami, Toshiya and Yamazaki, Hirohito and Iwamoto, Nobuyuki and Itoga, Toshiro and Kirihara, Yoichi and Lee, Eunji and Sugihara, Kenta and Miyamoto, Shuji and Hashimoto, Satoshi and Asano, Yoshihiro},
  year = 2025,
  month = nov,
  journal = {Physics Letters B},
  volume = {870},
  pages = {139900},
  issn = {03702693},
  urldate = {2025-10-21},
  url = {https://doi.org/10.1016/j.physletb.2025.139900}
}

@article{tomsPhotoprotonsCeBi1953,
  title = {Photoprotons from {{In}}, {{Ce}}, and {{Bi}}},
  author = {Toms, M. E. and Stephens, W. E.},
  year = 1953,
  month = oct,
  journal = {Phys. Rev.},
  volume = {92},
  number = {2},
  pages = {362--366},
  issn = {0031-899X},
  urldate = {2025-10-31},
  copyright = {http://link.aps.org/licenses/aps-default-license},
  month = {Oct},
  publisher = {American Physical Society},
  doi = {10.1103/PhysRev.92.362},
  url = {https://link.aps.org/doi/10.1103/PhysRev.92.362}
}

@article{tomsPhotoprotonsLead208Tantalum1955,
  title = {Photoprotons from {{Lead-208}} and {{Tantalum}}},
  author = {Toms, M. Elaine and Stephens, William E.},
  year = 1955,
  month = may,
  journal = {Physical Review},
  volume = {98},
  number = {3},
  pages = {626--628},
  issn = {0031-899X},
  urldate = {2025-10-31},
  copyright = {http://link.aps.org/licenses/aps-default-license},
  doi = {10.1103/PhysRev.98.626},
  url = {https://link.aps.org/doi/10.1103/PhysRev.98.626}
}

@article{VEYSSIERE1970561,
  title = {Photoneutron Cross Sections of {$^{208}$Pb} and {$^{197}$Au}},
  author = {Veyssiere, A. and Beil, H. and Bergere, R. and Carlos, P. and Lepretre, A.},
  year = 1970,
  journal = {Nuclear Physics A},
  volume = {159},
  number = {2},
  pages = {561--576},
  issn = {0375-9474},
  url = {https://doi.org/10.1016/0375-9474(70)90727-X}
}

@article{wafflerRelativeEffectiveCross1948,
  title = {{Relative Effective Cross Sections for the (\$gamma\$,n) Process with Lithium Gamma-Radiation (Quantum Energy = 17.5 Mev; Relative Wirkungsquerschnitte fur den (\$gamma\$,n)-Prozess mit der Lithium-Gamma-Strahlung (Quanten energie = 17.5 Mev}},
  author       = {Waffler, H and Hirzel, O},
  title        = {Relative Effective Cross Sections for the ($gamma$,n) Process with Lithium Gamma-Radiation (Quantum Energy = 17.5 Mev; Relative Wirkungsquerschnitte fur den ($gamma$,n)-Prozess mit der Lithium-Gamma-Strahlung (Quanten energie = 17.5 Mev},
  url          = {https://www.osti.gov/biblio/4438779},
  journal      = {HElvetica Physica Acta},
  volume       = {21},
  place        = {Country unknown/Code not available},
  year         = {1948},
  month        = {01}
}

@article{wilkinsonNUCLEARPHOTODISINTEGRATION1956,
  title = {{{NUCLEAR PHOTODISINTEGRATION}}},
  author = {Wilkinson, D. H.},
  year = 1956,
  journal = {Physica},
  volume = {22},
  pages = {1039},
  url = {https://doi.org/10.1016/S0031-8914(56)90061-1.}
}

@article{zatsepinaANGULARENERGYDISTRIBUTIONS1963,
  title = {{{ANGULAR AND ENERGY DISTRIBUTIONS OF PHOTONEUTRONS FROM BISMUTH}}, {{GOLD}}, {{AND TANTALUM}}},
  author = {Zatsepina, G N and Igonin, V V and Lazareva, L E},
  year = 1963,
  journal = {SOVIET PHYSICS JETP},
  volume = {17},
  langid = {english}
}
